# Mapping the magnonic landscape in patterned magnetic structures

C. S. Davies,[1,2] V. D. Poimanov,[3] and V. V. Kruglyak[1,*]

[1] *School of Physics, University of Exeter, Stocker road, Exeter, EX4 4QL, United Kingdom*

[2] *Radboud University, Institute for Molecules and Materials, 135 Heyendaalseweg, 6525 AJ Nijmegen, The Netherlands*

[3] *Faculty of Physics and Technology, Donetsk National University, 24 Universitetskaya Street, Donetsk 83001, Ukraine*

## Abstract

We report the development of a hybrid numerical / analytical model capable of mapping the spatially-varying distributions of the local ferromagnetic resonance (FMR) frequency and dynamic magnetic susceptibility in a wide class of patterned and compositionally modulated magnetic structures. Starting from the numerically simulated static micromagnetic state, the magnetization is deliberately deflected orthogonally to its equilibrium orientation, and the magnetic fields generated in response to this deflection are evaluated using micromagnetic software. This allows us to calculate the elements of the effective demagnetizing tensor, which are then used within a linear analytical formalism to map the local FMR frequency and dynamic magnetic susceptibility. To illustrate the typical results that one can obtain using this model, we analyze three micromagnetic systems boasting non-uniformity in either one or two dimensions, and successfully explain the spin-wave emission observed in each case, demonstrating the ubiquitous nature of the Schlömann excitation mechanism underpinning the observations. Finally, the developed model of local FMR frequency could be used to explain how spin waves could be confined and steered using magnetic non-uniformities of various origins, rendering it a powerful tool for the mapping of the graded magnonic index in magnonics.

---

[*] Corresponding author: V.V.Kruglyak@exeter.ac.uk

I. Introduction

The observed overwhelming progress in investigations of spin waves – precessional excitations of ordered spins in magnetic materials[1,2] – has established magnonics[3-5] as a compulsory element of the research roadmaps for magnetism.[6,7] Furthermore, several excellent recent reviews have highlighted interdisciplinary aspects of magnonics research, including (but not limited to) magnonic crystals and metamaterials,[8-13] photo-magnonics,[14] spin caloritronics,[15] magnon spintronics,[16-18] as well as applications of spin waves in computing and microwave signal processing.[19-21] In particular, the concept of *graded-index magnonics* has been proposed as a unifying theme focusing on general aspects of spin-wave excitation and propagation in media with continuously non-uniform properties.[22,23] This theme draws analogies with and feeds from other sub-fields of the wave physics (such as quantum mechanics,[24] graded-index optics[25] and transformation optics[26]), while also (and importantly) highlights phenomena resulting from aspects that are unique to the dispersion of spin waves, notably including the dispersion's nonlinearity, anisotropy and non-reciprocity.[1,2] Both the magnonic dispersion's complexity and the variety of factors that could be used to manipulate its character represent the main challenge and, at the same time, attraction of the graded-index magnonics for wave-physics experts. The factors affecting the spin-wave dispersion are best illustrated in the context of typical problems of the graded-index magnonics, such as spin-wave scattering, confinement, steering and emission.[23]

The compositional modulation of magnetic media represents the most obvious albeit technologically challenging way to modify the spin-wave dispersion.[27-29] Furthermore, any modification of surface magnetic properties could modify the surface boundary conditions and therefore the distribution of the spin-wave amplitude in the film thickness.[30-33] The dispersion of magnetostatic spin waves (i.e. those for which the effect of the exchange interaction is negligible) in thin-film magnetic samples is particularly sensitive to the variation of the film thickness.[34,35] Moreover, the finite thickness of magnetic film structures leads to quantization and thereby appearance of several dispersion branches for spin waves (of any sort) propagating in the film plane.[36,37] The same is true for the effect of the lateral quantization in waveguides of finite width.[38,39] The dispersion character is different for each branch and depends on the film thickness, which suggests the continuous variation of the thickness and / or width as a means through which to control the spin wave propagation in patterned magnetic structures,[40,41] and more generally, the importance of the sample geometry in problems of the graded-index magnonics. Magnonic crystals featuring the periodic modulation of the film width / thickness were reported in Refs. 42-44, while collective spin-wave modes in arrays of patterned magnetic elements were observed in Refs. 45-47.

Another aspect of the structure geometry is that, with the exception of ellipsoidal elements, the presence of lateral boundaries (or more generally, sharp non-uniformity of the magnetization length[48] or film thickness[49,50]) leads to a non-uniform demagnetizing field and therefore the internal magnetic field in their vicinity. The non-uniform internal magnetic field can lead to a non-uniform configuration of the magnetization,[22,42,45,47,51,52] while the magnetization is almost necessarily non-uniform in samples with spatially varying directions of the easy magnetization axes and / or a significant antisymmetric exchange interaction (Dzyaloshinskii-Moriya interaction).[53-58] The static magnetization and effective magnetic field are generally recognized as representing a landscape across which spin waves propagate.[12,22,23,40,42,45-60] The spatial variation of either the static magnetization and / or effective magnetic field can lead to spin-wave confinement,[61-66] while recently it has also been demonstrated to steer the direction of spin-wave propagation.[22,41,48-50,57,58,61,67,68] The scattering of spin waves from the non-uniformities of the internal field and magnetization were studied e.g. in Refs. 49,50,58,69,70. Magnetic non-uniformities can also be created by locally applied magnetic[20,59,71] and electric[72] fields, or optically,[68,73-75] offering an opportunity to study magnonic phenomena in time-varying graded-index magnonic landscapes.

Of importance for the present paper, the spatial non-uniformity of any kind (i.e. compositional, geometrical or micromagnetic) also opens up an alternative pathway for the excitation of propagating spin waves by microwaves. Most generally, the non-uniformity breaks the translational symmetry in the system, thereby enabling coupling between the incident microwave magnetic field and spin waves irrespective of their wavelengths. More specifically, pioneered by Schlömann for the case of a non-uniform applied magnetic field,[76,77] this mechanism of spin-wave emission exploits the fact that the spin-wave wavelength diverges at classical turning points and therefore matches the effectively infinite (compared to the dimensions of typical magnetic samples) wavelength of microwaves. The spin-wave emission from compositional magnetic non-uniformities was studied in Refs. 78,79. Geometry-enabled coupling of free-space microwaves to spin waves was studied in a series of our own earlier works,[22,43,80,81] with the importance of the dynamic non-uniform demagnetizing field highlighted in Refs. 82,83. The spin-wave emission from non-uniformities of the magnetization (such as magnetic domain walls and vortices) driven by microwaves was shown in Refs. 84,85.

The frequency at which a given point of the sample with position vector **r** serves as a turning point for spin waves may be called local ferromagnetic resonance (FMR) frequency, $\omega_0(\mathbf{r})$, i.e. the frequency of spin waves of infinite wavelength. Since the wavelength and the wave vector are not well defined in a medium lacking translational invariance, they need to be understood as in the Wentzel-Kramers-Brillouin (WKB) approximation.[24] Thus, for a given frequency $\omega$, the turning points (or lines in the two-dimensional and surfaces in three-dimensional cases) separating the evanescent and oscillatory solutions at either side are given by equation $\omega_0(\mathbf{r}) = \omega$. The oscillatory solutions may either form standing waves confined between two turning points, or alternatively have a propagating character. In the latter case, the turning point serves as a microwave-to-spin-wave transducer (antenna), i.e. a source of propagating spin waves.

An alternative interpretation of the Schlömann excitation mechanism is based on the notion of the local dynamic susceptibility $\hat{\chi}(\omega,\mathbf{r})$ tensor, which describes the strength of the magnetic medium's linear response at point **r** to excitation by a uniform microwave magnetic field of frequency $\omega$ in the approximation of absence of spatial dispersion.[86-88] The latter approximation means that the spatial derivatives are neglected in the calculation. Due to the requirement of the wave number (wavelength) matching, the frequency dependence of $\hat{\chi}(\omega,\mathbf{r})$ has a resonant character, peaking when the incident frequency matches the local FMR frequency. Thus, for a given microwave frequency, the incident microwave magnetic field is tuned to some regions of the graded magnonic medium better than to others. In particular, one could tune the microwave frequency so as to excite certain regions of the medium at resonance. The resonantly-driven magnetization precession launches spin waves of finite wave vector into the adjacent regions, if such propagating spin-waves are at all allowed by the dispersion relation in those regions.

The discussion above suggests the local FMR frequency as an important ingredient of the unified description of magnonic landscapes, i.e. graded magnonic index.[23] Indeed, considering $\omega_0(\mathbf{r})$ as the frequency of spin waves with zero wave vector and assuming that the character of the dispersion (i.e. its slope) is modified only weakly, the variation of $\omega_0(\mathbf{r})$ describes the displacement of the entire dispersion curve along the frequency axis. So, the role of the local FMR frequency for spin waves is similar to that of the potential for a quantum-mechanical electron.

In this paper, we report an approximate theoretical formalism that enables the mapping of the magnonic landscape in terms of the local FMR frequency $\omega_0(\mathbf{r})$, and give examples of its application to the problem of spin-wave emission in thin-film magnetic structures. The formalism is based on the effective demagnetizing factors,[86-88] which can be calculated using Object Oriented Micromagnetic Framework (OOMMF),[89] used here, or any other software for micromagnetic simulations. This formalism underpins our interpretation of the numerically and / or experimentally observed emission of spin waves from edges of patterned Permalloy and yttrium-iron-garnet (YIG) structures studied in Refs. 82,83,90. Here, the treatment is systematically generalized so as to show the full power of the approach. Indeed, we believe our formalism could explain a whole host of experimental results, since magnetic non-uniformities are inescapably present in magnetic systems of finite size. In addition, we believe that the local FMR frequency together with the effective demagnetizing factors and local dynamic susceptibility should find application in other problems of the graded-index magnonics, which are discussed here at a qualitative level only.

The structure of this report is as follows. In Section II, we describe the theoretical aspects of the model and give equations for the local FMR frequency and dynamic susceptibility. In Section III, we outline the numerical recipe used to obtain the components of the effective demagnetizing tensor across a micromagnetic system. These quantities underpin the utilization of the equations derived in Section II. In Section IV, we illustrate the validity, general nature, and usefulness of our formalism through investigation of three exemplary magnetic systems. The first system consists of an in-plane magnetized stripe with rectangular cross-section and infinite length. Due to the stripe being infinitely long, the magnetic non-uniformity is one-dimensional (i.e. varying only along the stripe's width). We demonstrate that our model can fully explain, on a quantitative basis, both the confinement of spin-waves ("edge-modes") and the possibility of exciting propagating spin waves due to the non-uniform dynamic demagnetizing field. The second exemplary system consists of an isolated antidot, i.e. hole patterned within an otherwise-continuous film. This system boasts magnetic non-uniformity in two dimensions, concentrated in the vicinity of the antidot. Our model enables the mapping of the local FMR frequency and magnetic susceptibility across a two-dimensional system, therefore explaining the observed excitation of spin-wave beams from the magnetization adjacent to the antidot. The third and final exemplary system is a circular disc in the vortex state. Our calculations predict and explain the potential for radially-coherent spin waves to be excited from both the edges of the disc (propagating inwards) and from the center of the disc (propagating outwards). This could offer a simple explanation for the experimental results reported in Ref. 85. In Section V, we discuss the limits of applicability of the presented formalism and list opportunities for its further development and exploitation. We then conclude this report with a summary of our findings.

## II. Theoretical formalism

In the continuous medium approximation, the dynamics of the magnetization $\mathbf{M}(\mathbf{r},t)$, and specifically excitation and propagation of spin waves, are described by the Landau-Lifshits-Gilbert equation[91,92]

$$\frac{\partial \mathbf{M}}{\partial t} = -\gamma [\mathbf{M} \times \mathbf{H}_{\text{eff}}] + \frac{\alpha_G}{M}\left[\mathbf{M} \times \frac{\partial \mathbf{M}}{\partial t}\right], \qquad (1)$$

where $\gamma$ is the gyromagnetic ratio, $\alpha_G$ is the Gilbert damping parameter, $M$ is the saturation magnetization, and $\mathbf{H}_{\text{eff}}(\mathbf{r},t)$ is the effective magnetic field. Considering spin waves of angular frequency $\omega$, one usually represents the total magnetization and effective field as $\mathbf{M} = \mathbf{M}_0(\mathbf{r}) + \mathbf{m}(\omega,\mathbf{r})\exp(i\omega t)$ and $\mathbf{H}_{\text{eff}} = \mathbf{H}_0(\mathbf{r}) + \mathbf{h}(\omega,\mathbf{r})\exp(i\omega t)$, where $\mathbf{h}(\omega,\mathbf{r})\exp(i\omega t)$ is the dynamic effective field (which can include the incident microwave magnetic field) and $\mathbf{m}(\omega,\mathbf{r})\exp(i\omega t)$ is the dynamic magnetization. The static effective magnetic field $\mathbf{H}_0$ and magnetization $\mathbf{M}_0$ satisfy the condition $[\mathbf{M}_0 \times \mathbf{H}_0] = 0$ in equilibrium. The static field may include contributions from the applied bias magnetic field $\mathbf{H}_B$, the static demagnetising field $\mathbf{H}$ (calculated from $\text{rot}\,\mathbf{H} = 0$, $\text{div}(\mathbf{H} + 4\pi\mathbf{M}_0) = 0$), magnetic anisotropies (of various symmetries and origin), the exchange interaction, as well as any other micromagnetic fields.

Assuming that the static problem has been solved and focusing on the small amplitude spin-wave excitations of the static micromagnetic configuration, one keeps in Equation (1) only terms linear in $\mathbf{m}$ or $\mathbf{h}$, so as to obtain the linearized Landau-Lifshits-Gilbert equation, which we write in the following form

$$i\omega M \mathbf{m} = -\gamma M [\mathbf{m} \times \mathbf{H}_0] - \gamma M [\mathbf{M}_0 \times \mathbf{h}] + i\omega \alpha_G [\mathbf{M}_0 \times \mathbf{m}]. \qquad (2)$$

Let us represent the dynamic effective field as a sum of the incident microwave field $\mathbf{h}_{\text{mw}}$ as well as the local $\mathbf{h}_{\text{loc}}$ and nonlocal $\mathbf{h}_{\text{nl}}$ contributions, i.e. $\mathbf{h} = \mathbf{h}_{\text{loc}} + \mathbf{h}_{\text{nl}} + \mathbf{h}_{\text{mw}}$. By definition, the local contribution $\mathbf{h}_{\text{loc}}$ depends on the dynamic magnetization at the same point in space and is therefore just an algebraic function of **m**. In contrast, the non-local contribution $\mathbf{h}_{nl}$ is generally related to **m** via a nonlocal operator, which may contain spatial derivatives of **m** (and / or even its integrals over the volume of the specimen). Hence, for $\mathbf{h}_{mw} = 0$, we may treat Equation (2) as a system of homogeneous algebraic equations that relate components of **m** and $\mathbf{h}_{nl}$. Then, the spatial variation of **m** emerges as a result of solving the system of differential equations determined by the definition of the non-local effective field in the problem and depending on the sample geometry and magnonic "landscape" given by spatially-dependent magnetic parameters with appropriate boundary conditions. Analytical solutions of the problem for graded magnonic landscapes are relatively rare and often obtained under strong approximations.[41,48,53,55,60-63,66,67,76,77,79] When solved numerically on a discrete mesh, the problem can be reduced to that of the dynamical matrix method (DMM), which can yield all the normal mode frequencies and profiles of the system in question[93,94] and its response to an external excitation (i.e. magnetic susceptibility).[87,95] However, the results of both DMM calculations and more conventional time-domain micromagnetic simulations are not readily interpreted in terms of the geometry, magnetic and micromagnetic landscapes of the studied system. Most commonly, the interpretation relies on the static effective field profile.[22,27,40,45,50,51,61-66,69,76-78,96,97] However, it is well-known that the dynamic effective field is at least as important in defining the frequencies of spin-wave normal modes, including the FMR frequency.[1,2,83,86,87] Hence, the definition of the magnonic landscape for the processes of spin-wave excitation and propagation should include contributions from both static and dynamic effective fields. Here, we propose that such a magnonic landscape can be defined in terms of the local FMR frequency $\omega_0(\mathbf{r})$.

We calculate the local FMR frequency $\omega_0(\mathbf{r})$ as follows. Firstly, we introduce the dynamical susceptibility tensor $\hat{\chi}(\omega,\mathbf{r})$ as

$$\mathbf{m} = \hat{\chi}(\omega,\mathbf{r})\mathbf{h}_{mw} \ , \qquad (3)$$

the spatial dependence of which is dictated not only by the static magnetic configuration, i.e. $\mathbf{H}_0(\mathbf{r})$ and $\mathbf{M}_0(\mathbf{r})$, but also by those contributions to the nonlocal field $\mathbf{h}_{nl}$ that remain non-zero in the limit of infinite spin-wave wavelength (i.e. uniform precession). The latter nonlocal contributions may include, for instance, the dynamic demagnetizing field[1,2,86] and / or curvature induced exchange anisotropy.[53,98,99]

Secondly, we introduce the effective demagnetizing tensor $\overleftrightarrow{N}$ such that

$$\mathbf{H}_0 = \mathbf{H}_B - 4\pi \overleftrightarrow{N}(\mathbf{r})\mathbf{M}_0, \tag{4}$$

$$\mathbf{H}_{\text{eff}} - \mathbf{H}_0 = -4\pi \overleftrightarrow{N}(\mathbf{r})(\mathbf{M} - \mathbf{M}_0), \tag{5}$$

where the effective magnetic field and $\overleftrightarrow{N}$ both include contributions from not only the real demagnetizing field but also the other micromagnetic energies (with exception of the Zeeman energy due to $\mathbf{H}_0$). The crucial assumption of our approach is that $\overleftrightarrow{N}(\mathbf{r})$ may be evaluated reasonably accurately from the modification of the effective magnetic field, $(\mathbf{H}_{\text{eff}} - \mathbf{H}_0)$, caused by an (approximately) uniform perturbation of the magnetization, $(\mathbf{M} - \mathbf{M}_0)$. Both the magnetization perturbation and the associated change of the effective field are defined in a special local curvilinear coordinate system, one axis of which (chosen to be *z*-axis here) is parallel to $\mathbf{M}_0$ everywhere. The magnetization perturbation is orthogonal to $\mathbf{M}_0$ and therefore lies in the *x-y* plane in this local coordinate system. The validity of this assumption is not guaranteed and forms the main topic of discussion in this report.

Using Equations (4-5) with Equation (2), and solving for each $\mathbf{r}$ the system of homogeneous algebraic equations obtained for $\mathbf{h}_{\text{mw}} = 0$, the local FMR frequency distribution is obtained in the form of a Kittel-like formula[100]

$$\omega_0(\mathbf{r}) = \gamma\sqrt{(H_{B,z} + 4\pi[N_{xx} - N_{zz}]M)(H_{B,z} + 4\pi[N_{yy} - N_{zz}]M) - (4\pi M)^2 N_{xy}N_{yx}}. \tag{6}$$

This shows that the spatial variation of the local FMR frequency can result from three sources: (i) the bias magnetic field itself, (ii) the magnetization's direction (as this influences the projection of the bias field onto the magnetization), and (iii) the effective demagnetizing factors. The two former sources as well as the spatial dependence of $N_{zz}$ are defined unambiguously. The same is true about the other components of $\overleftrightarrow{N}$ if they contain truly local contributions only. If, however, the nonlocal fields are present and the static magnetization configuration of a system is substantially non-uniform, it is generally impossible to define two orthogonal (relative to the static orientation) magnetization perturbations that would be both simultaneously uniform in the laboratory coordinate system. This makes the choice of the perturbations ambiguous and the corresponding values of the effective demagnetizing factors not uniquely defined. In practice, this requires one to manually define the local coordinate system and the magnetization perturbations so that the effective demagnetizing factors be close to the values characterizing (depending on the problem) either the quasi-uniform precessional mode of the system, or the driven precession of the system under the action of an external microwave field (which may be non-uniform).

This choice is further discussed in Section IV of this report, where specific applications of the method are presented.

For the case of precessional dynamics driven by an incident microwave magnetic field $h_{mw} \neq 0$, the non-zero components of the dynamical susceptibility tensor from Equation (3) are given by the standard equations

$$\chi_{xx,yy} = \frac{\gamma M\left(\omega_0^2 - \omega^2\right)\left(\omega_H + 4\pi\gamma M N_{yy,xx}\right) + i\left\{\alpha\omega\gamma M\left(\omega_0^2 - \omega^2\right) - 2\alpha\omega\omega_1\gamma M\left(\omega_H + 4\pi\gamma M N_{yy,xx}\right)\right\}}{\left(\omega_0^2 - \omega^2\right)^2 + (2\alpha\omega\omega_1)^2} \quad (7)$$

$$\chi_{xy,yx} = \frac{-4\pi(\gamma M)^2\left(\omega_0^2 - \omega^2\right)N_{xy,yx} \pm 2\gamma M \alpha\omega_1\omega^2 + i\left\{\pm\omega\gamma M\left(\omega_0^2 - \omega^2\right) + 2(4\pi)(\gamma M)^2 \alpha\omega_1\omega N_{xy,yx}\right\}}{\left(\omega_0^2 - \omega^2\right)^2 + (2\alpha\omega\omega_1)^2} \quad (8)$$

where $\omega_1 = \omega_H + \frac{1}{2}\gamma\left(N_{xx} + N_{yy} + N_{yx} - N_{xy}\right)\cdot 4\pi M$ and $\omega_H = \gamma(H_B - 4\pi N_{zz} M)$. This shows that, in addition to the sources identified for the local FMR frequency, the spatial variation of the dynamical susceptibility has a resonant dependence on the frequency of the incident microwave field, thereby enabling the Schlömann mechanism of spin-wave emission.

### III. Numerical recipe for evaluating the magnetic susceptibility tensor

To use the equations discussed in Section II, one needs to evaluate the elements of $\overleftrightarrow{N}$. At least in some cases, this can be done analytically.[86] However, the broadening use of numerical micromagnetic simulations in magnonics,[101] combined with the growing complexity of the systems being studied, generate a need for theoretical methods enabling the interpretation of the computed results in terms of the models' input specifications (and indeed physics) of the modeled spin-wave phenomena. Here, we describe how the local FMR frequency and dynamic susceptibility distributions can be calculated from the results of micromagnetic simulations performed using the OOMMF software,[89] although we consider the algorithm to be platform-independent.

In the first stage of our method, we numerically calculate the micromagnetic ground state $M_0(r)$ across the considered system and record all the constituent magnetic fields associated with the state, which are explicitly calculated by OOMMF. For simplicity, we consider in this report systems featuring the demagnetizing $H_d$ and exchange $H_{ex}$ fields only. Upon directing the z-axis of the local coordinate system locally with the orientation of the static magnetization, we can use the computed fields to immediately calculate $H_{B,z}$ and the $N_{zz}$ factor. Note that we assume that the magnetization dynamics stay within the linear approximation, and so no variation of the magnetization along the $z$-axis is allowed. So, to evaluate $N_{zz}$, we project all the static magnetic fields (except the bias field) onto the static magnetization, and use the definition

$$N_{zz} = -\frac{1}{4\pi M^2}\left[\left(\mathbf{H}_d + \mathbf{H}_{ex}\right)\cdot \mathbf{M}_0\right]. \tag{9}$$

To deduce the other demagnetizing factors, we now need to choose the $x$ and $y$ axes of the local coordinate system. In this report, we consider planar magnetic structures with mostly in-plane magnetizations. In this case, it is convenient to define the local coordinate system with the help of an auxiliary unit vector $\hat{\mathbf{n}}$ that is selected here so as to be normal to the sample's plane but could nominally point in any direction. The only requirement to the choice of $\hat{\mathbf{n}}$ is that one should avoid – as much as possible - the undefined situation where $\hat{\mathbf{n}}$ lies parallel to the magnetization vector in any point of the sample. Then, it is possible to define an axis that is orthogonal to both the static magnetization and $\hat{\mathbf{n}}$ with a unit vector

$$\hat{\mathbf{n}}_{IP}(x,y,z) = \frac{[\mathbf{M}_0 \times \hat{\mathbf{n}}]}{M}. \tag{10}$$

This axis is labelled as being "in-plane" (IP), due to it lying within the plane of the sample. The third axis, labelled "out-of-plane" (OOP), is then defined via

$$\hat{\mathbf{n}}_{OOP}(x,y,z) = \frac{[\hat{\mathbf{n}}_{IP} \times \mathbf{M}_0]}{M^2}. \tag{11}$$

This latter axis is orthogonal to both $\mathbf{M}_0$ and $\hat{\mathbf{n}}_{IP}$.

At the next stage of the data post-processing, we add to the calculated ground state magnetization $\mathbf{M}_0(\mathbf{r})$ modest deflections along $\hat{\mathbf{n}}_{IP}$ and $\hat{\mathbf{n}}_{OOP}$, such that

$$\mathbf{M}' = \mathbf{M}_0 + \frac{M}{1000}\hat{\mathbf{n}}_{IP} \quad \text{and} \quad \mathbf{M}'' = \mathbf{M}_0 + \frac{M}{1000}\hat{\mathbf{n}}_{OOP}. \tag{12}$$

Then, these "deflected" magnetization configurations are reloaded back into OOMMF, which computes at time zero the corresponding effective fields. The ground state magnetic fields and magnetization are then subtracted from their new values. Finally, we evaluate the relevant elements of the effective demagnetizing tensor from the equations

$$\begin{pmatrix} h'_{d,x} + h'_{ex,x} \\ h'_{d,y} + h'_{ex,y} \end{pmatrix} = -4\pi \begin{pmatrix} N_{xx} & N_{xy} \\ N_{yx} & N_{yy} \end{pmatrix}\begin{pmatrix} 0 \\ m'_y \end{pmatrix}, \tag{13}$$

when the magnetization was deflected along the $\hat{\mathbf{n}}_{\text{IP}}$ axis, and

$$\begin{pmatrix} h_{\text{d},x}'' + h_{\text{ex},x}'' \\ h_{\text{d},y}'' + h_{\text{ex},y}'' \end{pmatrix} = -4\pi \begin{pmatrix} N_{xx} & N_{xy} \\ N_{yx} & N_{yy} \end{pmatrix} \begin{pmatrix} m_x'' \\ 0 \end{pmatrix}, \quad (14)$$

when the magnetization deflection along the $\hat{\mathbf{n}}_{\text{OOP}}$ axis.

## IV    Discussion: Example applications of the theoretical model

The equations introduced and discussed in Section II allow for both the local FMR frequency and dynamic susceptibility to be mapped across a discretized magnetic configuration. To understand both the insights that one can expect from this model and its limitations, we study three test systems. The first system is a wide, infinitely-long stripe magnetized by an in-plane uniform bias magnetic field $\mathbf{H}_{\text{B}}$. Through analyzing the calculated distributions of $\omega_0(\mathbf{r})$ (or rather $f_0(\mathbf{r}) = \omega_0(\mathbf{r})/2\pi$) and $\hat{\chi}(\omega,\mathbf{r})$ as a function of both the thickness of the stripe and the orientation of $\mathbf{H}_{\text{B}}$, we reveal that our description is superior to / more insightful than the commonly-used description in terms of the static effective magnetic field. Since the non-uniformity of the magnetic configuration only spans across the width of the stripe, the distributions of $\omega_0(\mathbf{r})$ and $\hat{\chi}(\omega,\mathbf{r})$ are one-dimensional in this case. So, the second and third test systems are chosen to be non-uniform in two dimensions. Specifically, we consider the cases of an isolated antidot formed within an otherwise continuous magnetic film and a disk-shaped element hosting a magnetic vortex. All the test systems considered here are assumed to be composed of a Permalloy-like material, the gyromagnetic ratio, saturation magnetization, exchange constant, and Gilbert damping coefficient of which are assumed to be $\gamma / 2\pi = 2.8$ MHz / Oe, $M = 800$ G, $A_{\text{ex}} = 1.3$ μerg / cm, and $\alpha_{\text{G}} = 0.008$, respectively.

The three test systems have been studied either experimentally or numerically in literature,[82,83,85] where it has been shown that propagating spin waves can be excited in these systems by a uniform harmonic microwave magnetic field. Here, we demonstrate that our model is consistently able to explain the said generation of propagating spin waves. Moreover, we believe that our model is general enough to be able to predict and explain the emission of finite-wavelength spin waves from a wide class of magnetic non-uniformities, assuming that the latter can be appropriately described using micromagnetic simulations.

## A.  One-dimensional system - stripe

Fig. 1 (a) presents the geometry of the stripe that we consider. In the rest of this report, the laboratory and local coordinate systems are denoted with dashed and non-dashed variables. The stripe is assumed to have a fixed width $w = 40$ μm (along the $y´$-axis), while its thickness $s$ (along the $x´$-axis) is different in different simulations. The stripe's length (along the $z´$-axis) is fixed at 100 nm, but the one-dimensional periodic boundary conditions implemented along the $z´$-axis render the stripe infinitely long. Mesh cells of $(5 \times 5 \times s)$ nm$^3$ size were used to discretize the stripe. So, the in-plane cell sizes were about the exchange length of Permalloy (~5.7 nm). The bias magnetic field $\mathbf{H}_B$ had fixed strength of 500 Oe but was applied at different angles $\theta$ relative to the $z´$-axis, generally inducing spatially-varying rotation of the magnetization quantified by the angle $\alpha$.

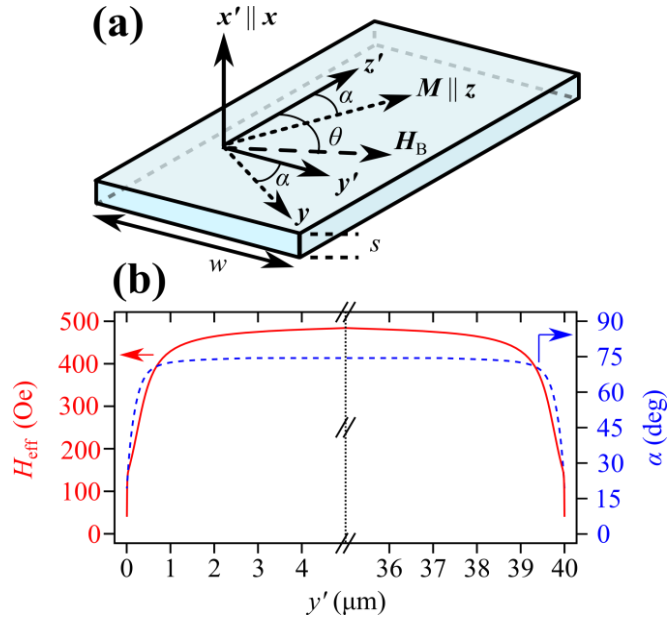

Fig. 1  (a) The considered stripe geometry is schematically shown together with the laboratory (dashed) and local (non-dashed) coordinate frames. The stripe is biased by magnetic field $\mathbf{H}_B$ of 500 Oe strength applied in-plane at an angle $\theta$ relative to the $z´$-axis (i.e. the stripe's long axis). The $z$-axis of the laboratory frame is parallel to the static magnetization, while its $y$-axis is orthogonal to both the $z$- and $x´$-axes. The $x$-axis is then orthogonal to both the $z$- and $y$-axes (and is parallel to the $x´$-axis, in this case). (b) The calculated spatial dependences of the projection of the static effective field onto the magnetization (solid red line) and the angle of the static magnetization $\alpha$ relative to the $z´$-axis (dashed blue line) close to the stripe's edges are shown for thickness $s = 40$ nm and for $\theta = 75°$.

Fig. 1 (b) shows the calculated spatial dependences of the projection of the effective magnetic field onto the static magnetization and the degree of canting of the static magnetization relative to the $z'$-axis. These distributions are calculated for the stripe's thickness $s = 40$ nm, and $\mathbf{H}_B$ applied at the angle $\theta = 75°$. Note that $H_{eff} = 491.4$ Oe and $\alpha = 75.0°$ at the center of the stripe ($y' = 20$ μm), changing only to $H_{eff} = 483.5$ Oe and $\alpha = 74.6°$ at $y' = 5$ μm and $y' = 35$ μm. So, the distributions of both $H_{eff}$ and $\alpha$ are reasonably uniform across the central 30 μm of the stripe and wholly symmetric about its center. This type of picture – involving the distributions of $H_{eff}$ and $\alpha$ – has been by far the most popular in the literature, e.g. illustrating well the existence of spin-wave edge modes[62] and providing basis for their more quantitative description.[63] However, this type of analysis cannot account for the excitation of propagating spin waves from the edge of a longitudinally magnetized stripe,[83] in which case the internal field and magnetization orientation both remain constant across the stripe's width.

Fig. 2 shows the results of our hybrid numerical / analytical formalism applied to the same sample and field geometry as described above. Let us discuss first the spatial variation of the demagnetizing factors shown in Fig. 2 (a). We neglect discussion of the off-diagonal elements, because these are negligibly small ($<1\times10^{-15}$) across the width of the stripe. The factors $N_{xx}$, $N_{yy}$ and $N_{zz}$ are uniformly equal to unity, zero and zero, respectively, across the central 30 μm across the stripe. This is due to the edges of the stripe being sufficiently far away, rendering the central 30 μm of the stripe to act alike a continuous thin film. However, a significant variation of the factors is observed closer to the edges. The factor $N_{yy}$ smoothly and monotonically increases towards a value of 0.5, when $y' < 0.5$ μm and $y' > 39.5$ μm. In a similar fashion, the factor $N_{xx}$ (not shown) decreases from unity towards a value of 0.5. The element $N_{zz}$ has a relatively small value, but a closer inspection (presented in the inset of Fig. 2 (a)) reveals that it does possess some variation.

Upon feeding the calculated demagnetizing tensor elements into Equation (6), we obtain the distribution $f_0(\mathbf{r})$ shown in Fig. 2 (b) alongside that of the effective magnetic field (reproduced from Fig. 1 (b)). Here, we clearly see that, as $y'$ increases, $f_0(\mathbf{r})$ first drops rapidly from a maximum of 14.0 GHz at the very edge of the stripe to a minimum of 5.0 GHz at $y' = 0.29$ μm and then further tends towards 6.4 GHz. This distribution of $f_0(\mathbf{r})$ can be understood as arising from the interplay between the static and dynamic demagnetizing fields close to the edge of the stripe. As one moves from the centre of the stripe towards the edge, the static demagnetizing field reduces the effective magnetic field, leading to a corresponding reduction of $f_0(\mathbf{r})$. However, at distances even closer to the edge, the dynamic demagnetizing field begins to grow rapidly, dominating the static demagnetizing field and resulting in the observed rise of $f_0$. This behavior is not an insignificantly small correction, since the rapid rise of $f_0$ occurs within 250 nm of the edge of the stripe (a distance that is easily resolvable using magneto-optical imaging techniques). Furthermore, it cannot be qualitatively deduced from the profile of the static effective magnetic field.

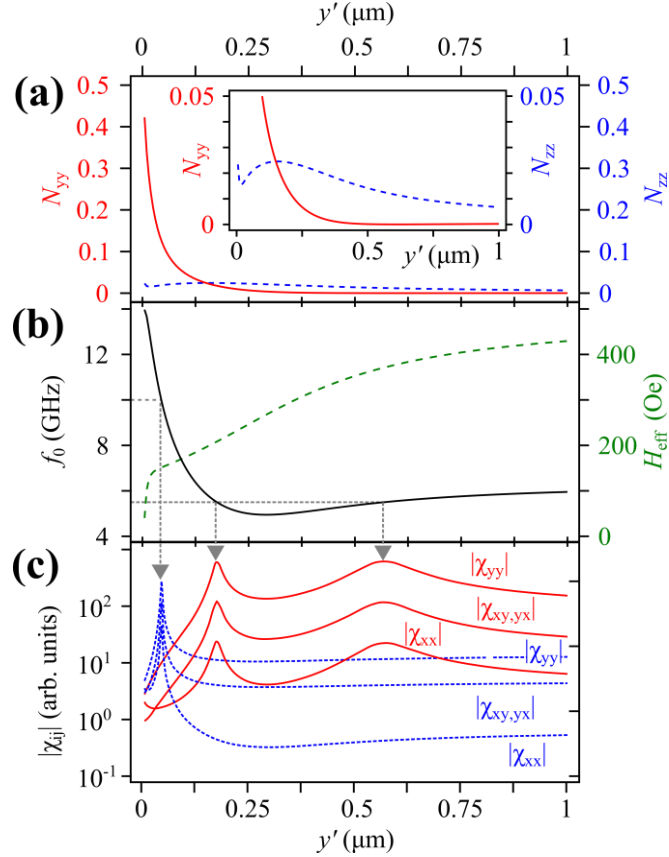

Fig. 2  (a) The distribution of the diagonal in-plane effective demagnetizing tensor elements $N_{yy}$ (red solid line) and $N_{zz}$ (blue dashed line) across the first micrometer from the stripe's edge. Inset: zoomed section of the distributions shown in the main panel. (b) The distribution of $f_0$ (black solid line) and the projection of the effective magnetic field onto the static magnetization (green dashed line) across the first micrometer from the stripe's edge. (c) The spatial variation of the elements of the susceptibility tensor as indicated, across the first micrometer from the stripe's edge. The red solid lines and blue dotted lines were calculated for the frequencies of 5.5 GHz and 10 GHz, respectively. The vertical scale has been scaled logarithmically for clarity. All profiles shown here are calculated for the system geometry from Fig. 1 (b), i.e. for a bias magnetic field of 500 Oe strength applied in-plane at an angle $\theta = 75°$ relative to the long axis of a 40 nm thick stripe.

Fig. 2 (c) shows the amplitude of the four elements of the susceptibility tensor calculated for the frequencies of 5.5 GHz and 10.0 GHz. For the frequency of 10 GHz, we observe one maximum at $y' = 45$ nm for all four elements, coinciding with the spatial coordinate that shares the same local FMR frequency. This demonstrates that the quantity $f_0$ corresponds to the maximum of the local dynamic susceptibility. The relative values of the susceptibilities consistently behave according to $|\chi_{yy}| > |\chi_{yx,xy}| > |\chi_{xx}|$, as expected. For the frequency of 5.5 GHz, in contrast, there are two positions of maxima, located at $y' = 175$ nm and $y' = 570$ nm. Again, these correspond to the same points in space that boast the corresponding value of $f_0 = 5.5$ GHz.

To directly confirm the validity of the above analysis, we excite the stripe using a spatially-uniform driving magnetic field **h** given by

$$\mathbf{h} = h_0 \sin(\omega t)\hat{\mathbf{x}}', \qquad (15)$$

where $h_0 = 0.1$ Oe is the amplitude of the excitation, and $\omega = 2\pi f$ where $f$ is the oscillation frequency. This excitation was deliberately applied solely along the $x'$-axis, so that this field exerts the same torque across all the magnetization vectors within the stripe. The magnetization across the stripe was then sampled in time steps of $T/32$, where $T = 1/f$. Fig. 3 (a) and (b) shows the time-varying profiles of the out-of-plane component of the dynamic magnetization close to the edge of the stripe excited at frequencies of 10 GHz and 12 GHz, respectively. In both cases, the magnetization precession is concentrated close to $y' = 0$ at small time, and as time increases, the precession spreads towards greater values of $y'$ in a phase-coherent manner, i.e. we observe spin-wave propagation away from the edge of the stripe. For the higher frequency, the wavelength is shorter and the group velocity is greater, as anticipated from the dipole-exchange spin-wave dispersion relation.[2] This set of observations cannot be qualitatively interpreted directly from the spatially-resolved profiles of the effective magnetic field and magnetization orientation as presented in Fig. 1 (b), but our description involving $f_0(\mathbf{r})$ and $\hat{\chi}(\omega,\mathbf{r})$ explains the observed behaviour completely.

An important point that pertains to the results presented in Fig. 3 relates to the magnetization oscillations that are observed in the background. When $t$ is small, one can observe beating in the background oscillations (for $y' > 1$ μm). This beating arises as follows. The magnetization far from the stripe's edges has a local FMR frequency that coincides with the quasi-uniform FMR frequency of the sample as a whole and is about half of the driving frequency. Nonetheless, the magnetization there still undergoes driven (off-resonance, and so, less efficiently) precession. This effect will (and indeed does) persist indefinitely, therefore representing a parasitic feature accompanying the discussed excitation mechanism. In addition, one needs to consider that the onset of the uniform driving field **h** in the simulations is abrupt in time, causing the magnetization to be excited at all frequencies rather than just the "carrier" one. Due to its uniformity of **h**, this broadband spectral content of the excitation couples most strongly to the quasi-uniform FMR mode (6.4 GHz), although the associated magnetization precession then decays quickly within ~2 ns.

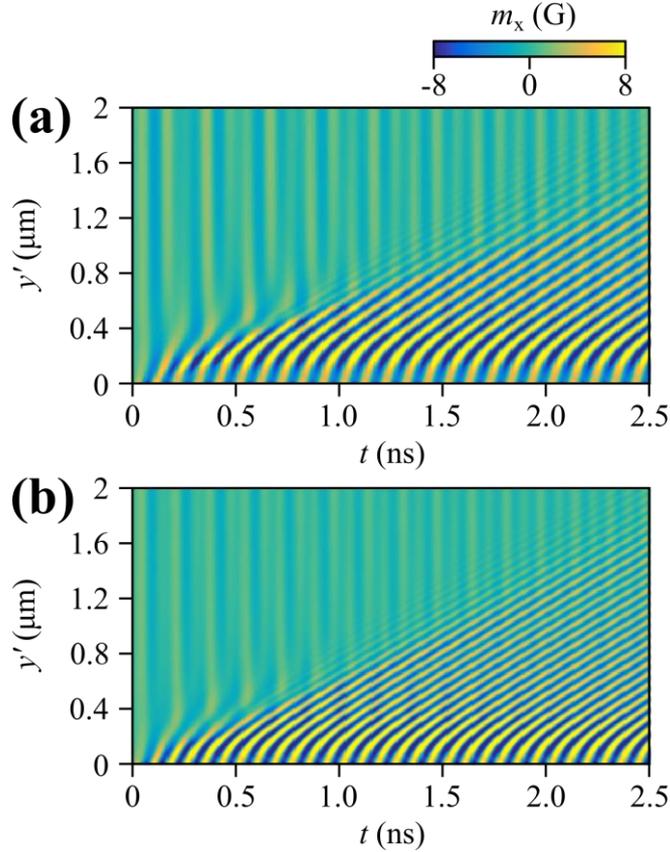

Fig. 3   Panels (a) and (b) show the simulated time evolution of the spin-wave profile close to edge of the 40 nm thick stripe driven by a uniform microwave field of 10 GHz and 12 GHz frequency, respectively. The uniform bias magnetic field of 500 Oe is applied at an angle of $\theta = 75°$ relative to the $z'$-axis. The amplitude of the driving magnetic field (directed along the $x'$-axis) is 0.1 Oe.

The orientation of the bias magnetic field substantially influences the distribution of both the local FMR frequency and the local dynamic susceptibility. So far, the stripe has been magnetized almost transversely, causing there to be competition between the static and dynamic demagnetizing fields close to the stripe edge. To investigate how this competition depends on $\theta$, we repeat the previously discussed calculations for different orientations of the bias magnetic field applied to the 40 nm thick stripe. In each case, the magnetization is initially uniformly aligned along the $z'$-axis and then allowed to relax for the given orientation of $\mathbf{H}_B$. Upon reducing $\theta$, the projection of the effective magnetic field onto the static magnetization (not shown) becomes increasingly flat across the stripe and approaches $H_{\text{eff}} = 500$ Oe, reflecting the reduction in the strength of the static demagnetizing field. Similarly, the magnetization (Fig. 4 (a)) aligns more uniformly with the long axis of the stripe. For the case of $\theta = 0°$, $H_{\text{eff}}(\mathbf{r}) = 500$ Oe and $\alpha(\mathbf{r}) = 0°$.

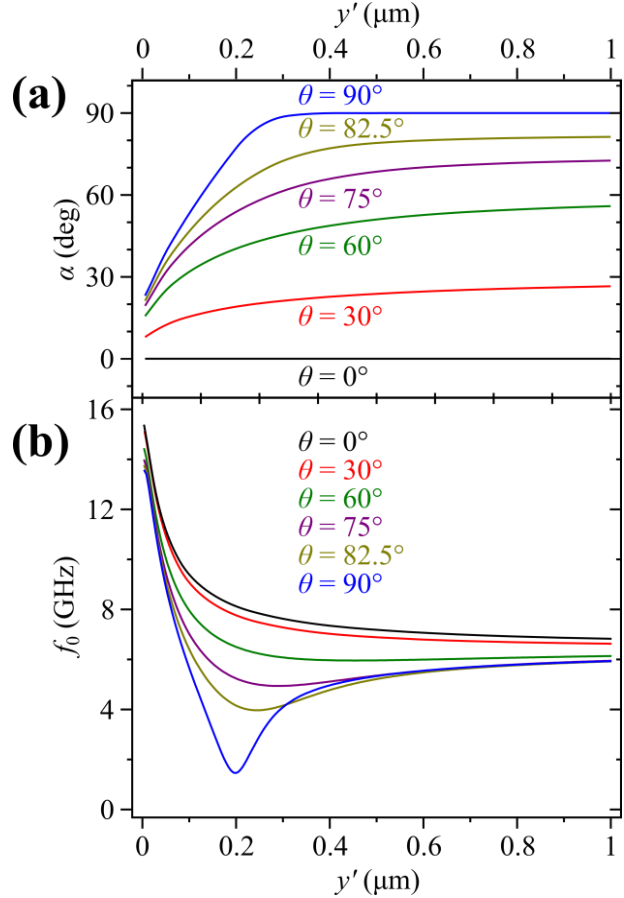

Fig. 4 (a) The calculated angle of canting between the static magnetization and the $z'$-axis and (b) the corresponding variation of $f_0$ are shown for the first micrometer from the edge of a 40 nm thick stripe to which the bias magnetic field of 500 Oe is applied at the indicated values of the angle $\theta$ relative to the $z'$-axis.

Expectedly, the variation of $\theta$ results in a significant deformation of the $f_0(\mathbf{r})$ profile (Fig. 4 (b)). When $\theta = 0°$, the contribution from static demagnetization is entirely eliminated, and only the dynamic demagnetization exists instead. At $y' = 0$, $f_0$ is maximised at 15.36 GHz, and as $y'$ increases, $f_0$ smoothly and monotonically decreases to a minimum of 6.48 GHz. When $\theta = 30°$, the variation of $f_0$ remains monotonic, but the reduction close to the stripe's edge is sharper. When $\theta = 60°$, $f_0$ dips slightly to 5.96 GHz at $y' = 455$ nm and then approaches $f_0 = 6.4$ GHz for higher $y'$. As $\theta$ increases further, the dip shifts closer to the edge of the stripe and becomes more pronounced, reaching a minimum of 1.48 GHz at $y' = 200$ nm when $\theta = 90°$.

The reduction in $f_0$ observed when $\theta = 90°$ is highly reminiscent of a potential well.[62] Upon exciting the entire stripe at a frequency within this "well", the driven magnetization precession is confined to the well (Fig. 5 (a)). This precession does not necessarily correspond to a confined spin-wave edge mode (which has a discrete spectrum) but the precession strength is maximised when the excitation frequency approaches that of the edge mode. In Fig. 5 (a), the driving magnetic field has frequency of 4 GHz. Fig. 5 (b), in contrast, shows the result of driving the stripe at 12 GHz frequency, irrefutably showing that propagating spin waves are emitted from the stripe's edge. These results demonstrate the stark difference between the excitation of the confined edge-modes[62-65] and the propagating spin waves identified in Ref. 83.

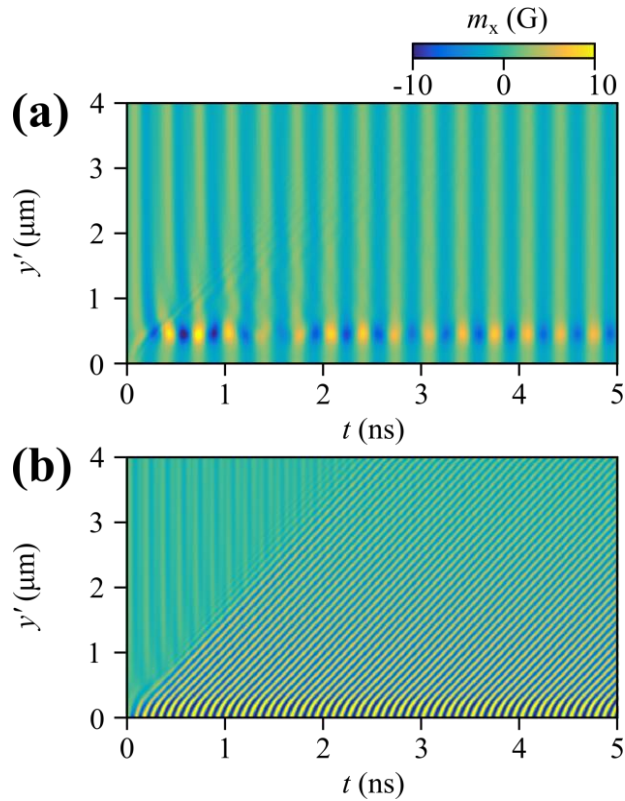

Fig. 5  Panels (a) and (b) show the simulated time evolution of the spin-wave profile close to edge of the 40 nm thick stripe driven by a uniform microwave field of 4 GHz and 12 GHz frequency, respectively. The uniform bias magnetic field of 500 Oe is applied along the $y'$-axis. The amplitude of the driving magnetic field (directed along the $x'$-axis) is 0.1 Oe.

As a final point of this section, we discuss the influence of the stripe thickness on the profile of $f_0(\mathbf{r})$. It is well understood that, as the magnetic film thickness increases, the demagnetizing field (either static or dynamic) originating from the film edge extends further from the edge. So, Fig. 6 (a) shows the distributions of $f_0$ calculated for longitudinally magnetized stripes with their thickness varied in increments of 35 nm. Indeed, we observe that, as the stripe thickness increases, the reduction in $f_0$ from the edge towards the centre of the stripe becomes more gradual. Focusing on the 40 nm thick stripe, Fig. 6 (b) shows the calculated distributions of both the local FMR frequency and the magnetic susceptibility for the frequencies of 7 GHz and 9 GHz. Again, we note that the maximum of the susceptibility occurs at the spatial coordinate sharing the matching value of $f_0$.

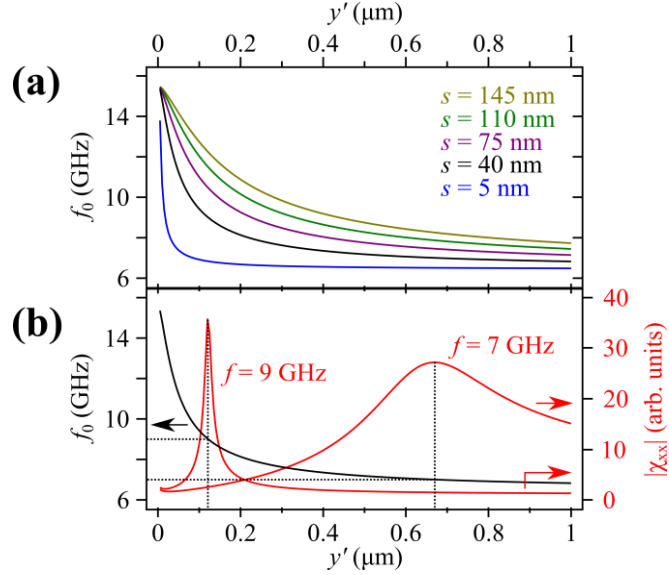

Fig. 6  (a) The calculated distributions of $f_0$ are shown for the first micrometer from the edge of stripes of varying thickness as indicated. The stripes are uniformly magnetized along their long axes by a bias magnetic field of 500 Oe. (b) The black line shows the distribution of $f_0$ for the 40 nm thick stripe from panel (a) as a reference for the corresponding susceptibility profiles shown by the red lines for the 9 GHz and 7 GHz frequencies.

B.  *Two-dimensional system – circular antidot*

In Refs. 82,83, we demonstrated that the Schlömann mechanism of spin-wave generation is active in the case of an isolated antidot resonantly excited by a uniform microwave magnetic field. Here, we use our calculations to deduce more precisely the spatial location of the magnetization responsible for the emission at a particular frequency. We consider a 6 nm thick magnetic film simulated by applying two-dimensional periodic boundary conditions (in the $y'$-$z'$ plane) to a square of a 6 μm side, at the center of which an antidot of 200 nm diameter is formed. A uniform bias magnetic field $H_B$ = 500 Oe is applied along the $z'$-axis. It is well understood[82,82] that the magnetic charges generated at the top and bottom of the antidot generate a static demagnetizing field that opposes the bias magnetic field above and below the antidot but actually strengthens it to the left and to the right of the antidot. Thus, it is expected that the local FMR frequency in the latter regions is boosted by both the static and dynamic demagnetizing fields.

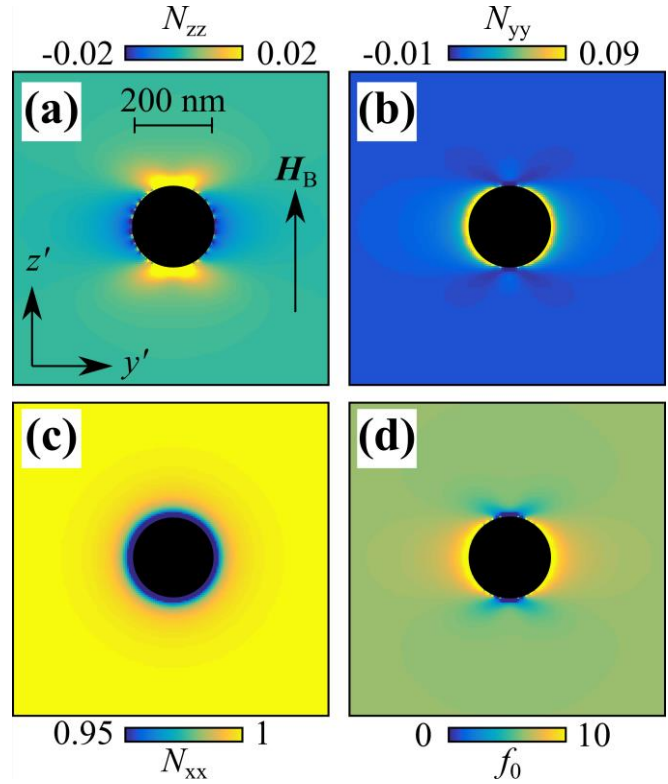

Fig. 7 (a) – (c) The spatial distributions of the effective demagnetizing factors $N_{zz}$, $N_{yy}$ and $N_{xx}$, respectively, are shown for an antidot of 200 nm diameter. (d) The corresponding distribution of the local FMR frequency is shown. The film is biased by a magnetic field of 500 Oe along the $z'$-axis.

Fig. 7 (a-c) shows the calculated spatial distributions of $N_{zz}$, $N_{yy}$ and $N_{xx}$, which illustrate a few important features of our theory. In the vicinity of the antidot, the effective demagnetizing factors vary significantly, while $N_{zz} = N_{yy} \sim 0$ and $N_{xx} \sim 1$ far from the antidot, as for a continuous film. The distribution of $N_{xx}$ is radially symmetric. This is expected for a circular antidot in a radially symmetric magnetization state. Indeed, $N_{xx}$ is calculated by giving the in-plane static magnetization a uniform out-of-plane deviation. The distributions of $N_{zz}$ and $N_{yy}$ have two-fold symmetry but are different. The difference is a result of the non-uniformity of the static magnetization, which leads to a non-equivalence of in-plane rotations that are uniform in the laboratory and local coordinate systems.

In some regions adjacent to the antidot edge, one of the in-plane demagnetizing factors, i.e. either $N_{zz}$ or $N_{yy}$, turns negative, which means that the corresponding components of the effective demagnetizing field and the magnetization used in the calculation have the same rather than opposite signs. This is physically permitted, except when the corresponding value of the local FMR frequency becomes negative. The spatial distribution of $f_0$ is shown in Fig. 7 (d). Consistent with the qualitative description given in Ref. 82, the local FMR frequency is about 6.4 GHz far from the antidot, which is equal to the FMR frequency of the continuous thin film, and is smaller than 6.4 GHz above and below the antidot, becoming negative just near its edge. This results from a failure of our approximation whereby we attempt to calculate the effective demagnetizing factors in response to uniform deviations of the magnetization from its static state. More generally, at distances close to the edge where the local FMR frequency is negative (mainly due to a strong negative static demagnetizing field), this indicates that one cannot neglect the spatial dispersion in such regions of the sample. So, all imaginary frequencies of $f_0$ are manually ascribed zero values in Fig. 7 (d).

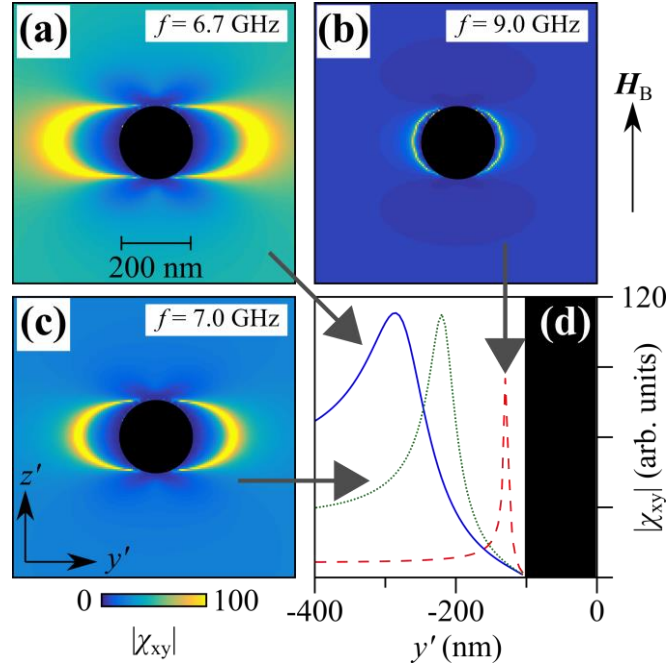

Fig. 8 (a)-(c) The calculated spatial variation of the absolute value of $\chi_{xy}$ in the vicinity of the antidot is shown for the frequencies of 6.7, 9 and 7 GHz, respectively. (d) The cross-sections of the susceptibility profiles from panels (a)-(c) through the antidot's equator are shown for the frequencies of 6.7 GHz (blue solid line), 7 GHz (green dotted line), and 9 GHz (red dashed line). $y´ = 0$ corresponds to the antidot's center.

In agreement with Ref. 82, the local FMR frequency exceeds the thin-film value of 6.4 GHz on the left- and right-hand sides of the antidot, with the maximum value of 16.4 GHz recorded at the very edge. This local enhancement of the local FMR frequency underpins the Schlömann mechanism of spin-wave generation. To verify this, the distributions of the local susceptibility $\chi_{xy}$ calculated for the frequencies of 6.7, 9 and 7 GHz are shown in Fig. 8 (a)-(c), respectively. As expected from the local FMR frequency distribution, we observe that, as the frequency increases, the maximum of the susceptibility distribution shifts closer to the antidot's edges and becomes sharper. At the next stage, we performed dynamic micromagnetic simulations, in a similar manner to those discussed in Ref. 83. Snapshots of the dynamic magnetization generated in response to the microwave magnetic field defined by Equation (15) are presented in Fig. 9. The magnetization close to the left- and right-hand sides of the antidot is excited most efficiently, causing the emission of spin waves propagating towards the left and right. One can also see the parasitic background oscillation, which emerges from both the driven magnetization precession and the abrupt onset of the excitation.

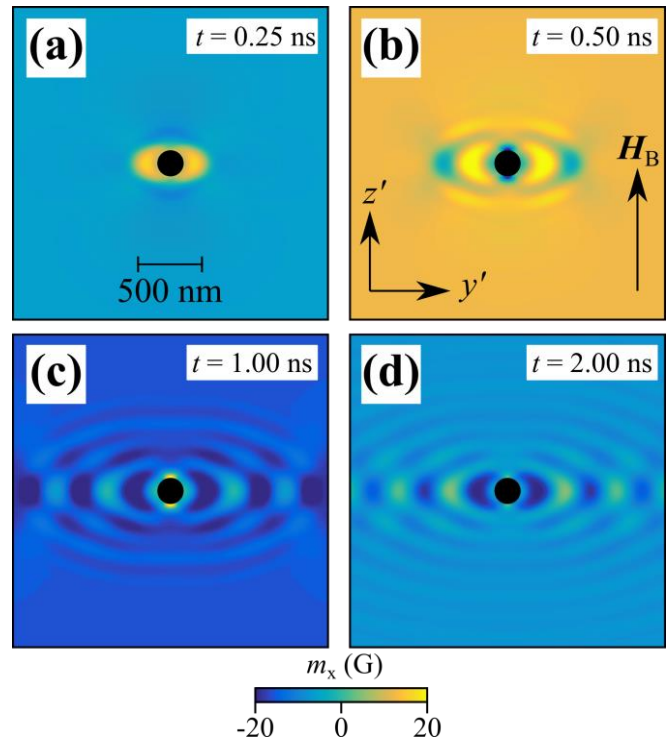

Fig. 9  (a)-(d) Snapshots of the dynamic out-of-plane component of magnetization near the antidot are shown for 0.25, 0.5, 1 and 2 ns, respectively, after the onset of an out-of-plane microwave magnetic field of 9 GHz frequency.

## C. Two-dimensional system – circular disk in a vortex state

Recently, Wintz *et al* used time- and layer-resolved scanning transmission X-ray microscopy to image the magnetization dynamics of two magnetic vortices hosted by an antiferromagnetically-coupled magnetic trilayer stack patterned into a disk of 4 μm diameter.[85] The acquired images revealed the excitation of radially-coherent spin-wave packets, emanating either inwards towards or outwards away from the vortex core. Here, we attempt to isolate the mechanism of spin-wave emission observed in Ref. 85 by considering a somewhat simpler system – an isolated disk of Permalloy with thickness of 30 nm (along the $x'$-axis) and diameter of 5 μm. No bias magnetic field is applied. Instead, the magnetization is allowed to form the vortex state, schematically shown in the inset of Fig. 10 (b).

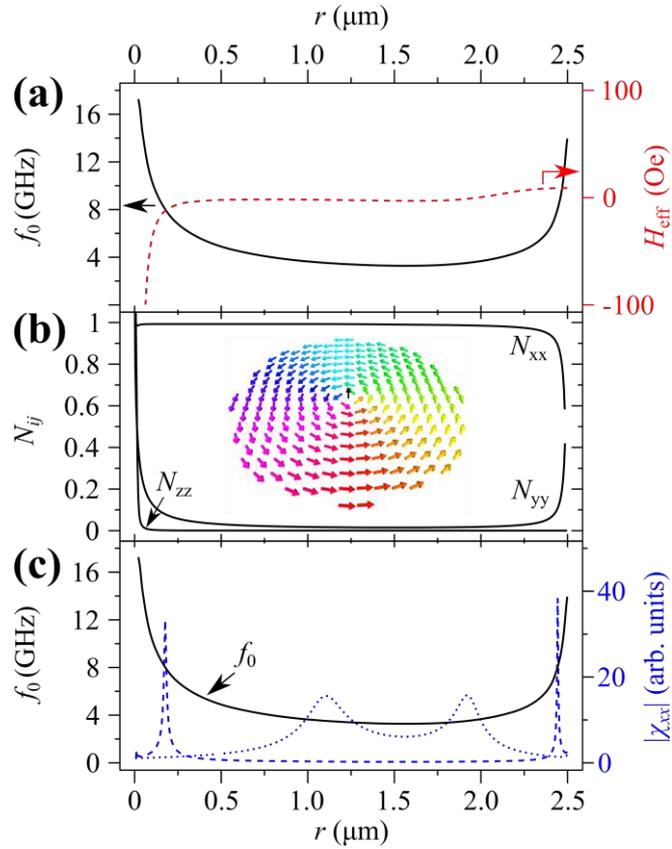

Fig. 10 The calculated distributions along the disk radius of (a) the projection of the effective magnetic field onto the static magnetization (dashed red line) and of $f_0$ (solid black line); (b) the diagonal elements of the demagnetizing tensor (inset: a schematic illustration of the magnetic vortex); (c) the absolute value of the local susceptibility element $\chi_{xx}$ for the frequencies 3.5 GHz (blue dotted line) and 8 GHz (blue dashed line), together with $f_0$ (black line) repeated from panel (a).

Fig. 10 (a) shows the calculated distribution of the projection of the effective magnetic field onto the static magnetization along the radius of the disk. Close to the disk edges ($r \sim 2.5$ μm), the effective magnetic field is very uniform and small. This regime persists until about $r \sim 0.25$ μm, after which the effective field drops rapidly at distances even closer to the vortex core. This behavior is in striking contrast to that of the calculated local FMR frequency distribution of $f_0(\mathbf{r})$, also shown in Fig. 10 (a). The local FMR frequency reaches value of 14 GHz close to the edge of the disk, dropping rapidly to 5 GHz at $r \sim 2.3$ μm and then minimizing at 3.3 GHz. Then, as $r$ decreases below 1.5 μm, the local FMR frequency increases again (more gradually) towards a maximum of 17.3 GHz.

The spatial variation of the diagonal elements of the demagnetizing tensor are shown in Fig. 10 (b). The factor $N_{zz}$ is zero across the majority of the disc, and increases only when dramatically close to the center. $N_{yy}$, in contrast, shows a characteristic enhancement close to the center and edges of the disk, which mimics the behavior of $f_0$, and is zero otherwise. Complementarily, $N_{xx}$ is unity across the majority of the disk and diminishes in the center and near the edges. Near the center of the disk, all the elements increase dramatically towards positive values in excess of unity. On one hand, this indicates the increased contribution from the exchange interaction to their values. On the other hand, as discussed earlier, the strongly non-uniform magnetization increasingly aligned with the disk normal in this region invalidates our main assumption that the spatial dispersion can be neglected in the calculation. We note that the corresponding pixel values have been removed from the distributions shown in Fig. 10. Fig. 10 (c) shows the radial distributions of the susceptibility calculated for the frequencies of 3.5 GHz and 8 GHz. These corroborate our earlier findings that the distribution of the local FMR frequency describes positions at which the local susceptibility peaks at a given value of the incident microwave frequency.

In the context of the experimental observations from Ref. 85, we note that the radially symmetric distribution of $f_0$ has regions of raised values close to both the center and edges of the disk. This suggests that, upon application of a uniform microwave magnetic field, counter-propagating radial spin waves could be simultaneously excited both from the center and edges of the disk. To verify this hypothesis, we perform a series of dynamic micromagnetic simulations, in which the disk in the vortex state was driven by a microwave magnetic field similar to that defined in Equation (15). Presented in Fig. 11 (a)-(c), from top to bottom, are snapshots of the dynamic magnetization in the disk, stimulated with the driving magnetic field of 12.5 GHz frequency directed along the in-plane horizontal, in-plane vertical and out-of-plane axes, respectively. In Fig. 11 (a), we observe that the spin waves with opposite phase are initiated close to the left and right edges of the disk and propagate inwards from the edges. In addition, we also observe some weaker oscillations propagating outwards from the center of the disk (again with opposite phase). After some time, a stationary interference pattern forms, with a characteristic spiraling structure. In Fig. 11 (b), a similar pattern is observed, except that the spin waves propagate along the vertical axis. The absence of spin-wave emission along the horizontal and vertical axes in Figs. 11 (a) and (b), respectively, is easily understood as arising from the torque exerted by the spatially-uniform ac magnetic field on the spatially-varying magnetization constituting the magnetic vortex state. Hence, upon removing this mismatch through applying an out-of-plane microwave magnetic field, we observe that the spin-waves excited from the edges and center of the magnetic disc are radially-symmetric. We note in passing that additional micromagnetic simulations (not shown) reveal that reversal of the chirality of the magnetic vortex flips the phase of the excited spin waves but does not affect their intensity). The core polarity has no identifiable impact on the dynamics triggered. The spectrum of standing spin waves in magnetic disks in the vortex state were extensively studied in a number of theoretical works.[102-105] The theoretical framework proposed here complement them by enabling insights into the mechanism of excitation of propagating spin-wave modes in such systems.

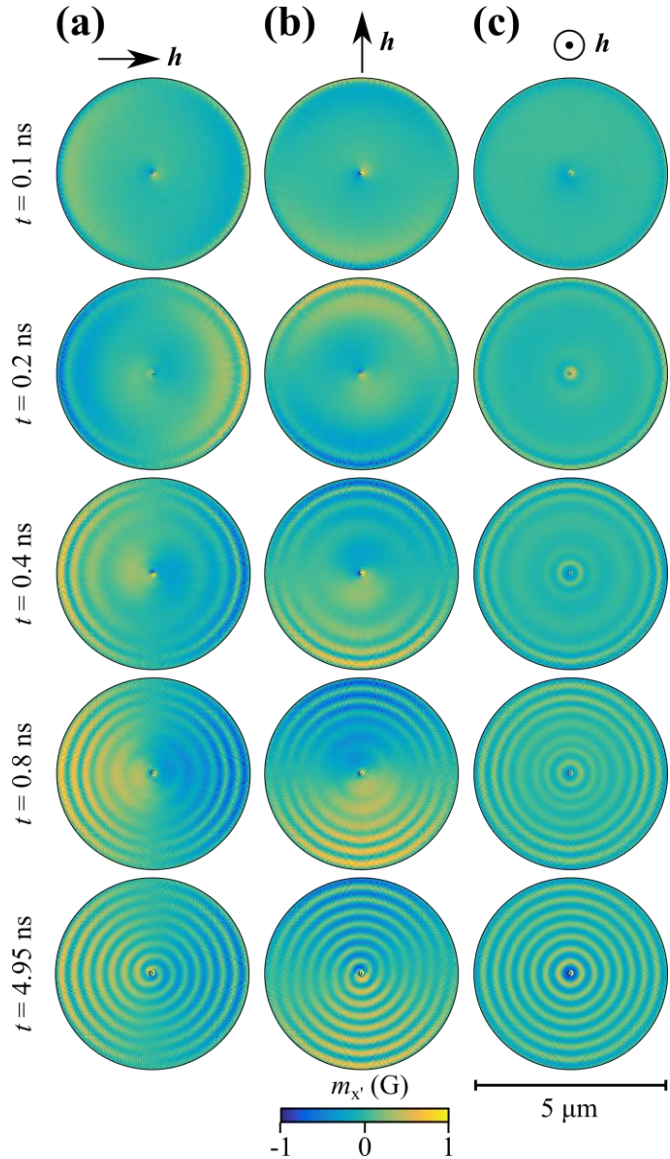

Fig. 11 (a)-(c) Simulated snapshots of the dynamic out-of-plane component of magnetization in the disk are presented from top to bottom for the indicated increasing time delays after the onset of the horizontal, vertical and out-of-plane driving magnetic field of 12.5 GHz frequency, respectively.

## V. Summary

All real patterned magnetic structures exhibit non-uniformity of some kind: e.g. through the static micromagnetic configuration, the dynamic demagnetizing field or through a deliberate design of compositional magnetic inhomogeneities. This non-uniformity naturally leads to the local FMR frequency and local dynamic susceptibility being spatially-varying. In this report, we have introduced and discussed a theoretical model that is capable of describing the graded variation of these quantities across non-uniform magnetic configurations. To illustrate the wide applicability of our model, we have studied three example magnetic systems that feature non-uniformity. Firstly, we have extensively studied the magnonic landscape associated with a stripe, and demonstrated that our model can quantitatively account not only for the static micromagnetic effects but also those associated with the dynamic demagnetizing field induced by the magnetization precession. Our dynamic micromagnetic simulations have revealed that, at low excitation frequencies, one can generate precession confined to the edge regions, and at high excitation frequencies, one can excite spin waves propagating away from the edge of the stripe. Our theoretical model can explain both these effects. Secondly, we have mapped the magnonic index in the vicinity of an isolated antidot and demonstrated quantitatively how the frequency of resonance varies in this region. Again, dynamic micromagnetic simulations have been performed to verify our findings. Thirdly, we have investigated how our model can be applied to study a magnetic disk that hosts a magnetic vortex state. Our results demonstrate that radially-coherent counter-propagating spin waves can be excited from both the outer edge and inner core of the magnetic vortex. We believe that, within the outlined approximations, our model can be used to map the graded magnonic index across a wide class of samples and micromagnetic landscapes and therefore will have significant impact on the theoretical understanding of spin-wave excitation mechanisms in general.

**Acknowledgements**

The research leading to these results has received funding from the Engineering and Physical Sciences Research Council of the United Kingdom (Project Nos. EP/L019876/1 and EP/P505526/1), and from the European Union's Horizon 2020 research and innovation program under Marie Skłodowska-Curie Grant Agreement No. 644348 (MagIC).